\documentclass{ifacconf}

\usepackage{natbib}   

\usepackage[pdftex]{graphicx}
\DeclareGraphicsExtensions{.pdf,.jpg,.png,.mps} 
\graphicspath{{plots/}}

\usepackage{units}
\usepackage{amsmath,amssymb}  
\usepackage[utf8]{inputenc}

\newcommand{\ve}[1]{\ensuremath{\mathbf{#1}}}
\newcommand{\set}[1]{\mathcal{#1}}
\newcommand{\op}[1]{\mathrm{#1}}
\newcommand{\transp}{\mathtt{T}}

\newcommand{\coordf}[0]{C}

\usepackage{pgf}

\begin{document}

\begin{frontmatter}

\title{Computing Feasible Vehicle Platooning Opportunities for Transport Assignments\thanksref{footnoteinfo}} 

\thanks[footnoteinfo]{This work was supported by the COMPANION EU project, the Knut and Alice Wallenberg Foundation, and the Swedish Research Council.}

\author[First]{Sebastian van de Hoef} 
\author[First]{Karl H. Johansson} 
\author[First]{Dimos V. Dimarogonas}

\address[First]{ACCESS Linnaeus Center and the School of Electrical Engineering, KTH Royal Institute of Technology, SE-100 44, Stockholm, Sweden (e-mail: {\tt \{shvdh, kallej, dimos\}@kth.se}).}

\begin{abstract}
Vehicle platooning facilitates the partial automation of vehicles and can significantly reduce fuel consumption. Mobile communication infrastructure makes it possible to dynamically coordinate the formation of platoons en route. We consider a centralized system that provides trucks with routes and speed profiles allowing them to dynamically form platoons during their journeys. For this to work, all possible pairs of vehicles that can platoon based on their location, destination, and other constraints have to be identified. The presented approach scales well to large vehicle fleets and realistic road networks by extracting features from the transport assignments of the vehicles and rules out a majority of possible pairs based on these features only. Merely a small number of remaining pairs are considered in depth by a complete and computationally expensive algorithm. This algorithm conclusively decides if platooning is possible for a pair based on the complete data associated with the two vehicles. We derive appropriate features for the problem and demonstrate the effectiveness of the approach in a simulation example.
\end{abstract}

\begin{keyword}
Transportation, Agents, Platooning, Spatial Networks
\end{keyword}

\end{frontmatter}

\section{Introduction}

Platooning is one of the fundamental building blocks for controlling connected vehicles. Vehicles are arranged in a convoy and the longitudinal spacing in the convoy is maintained with the help of automatic control. This simplifies the automation of the trailing vehicles and reduces their fuel consumption due to the slipstream effect (\cite{Bonnet2000}). While the low-level platoon control is well developed (\cite{path_overview_conference, barooah_platooning}), the dynamic formation of platoons has only recently attracted the interest of researchers (\cite{datamining_platooning, Hall_platoon_sorting, jeff_kuo_yun_distributed_controller, Larsson_Platoon_Complexity, ACCpaper, ITSCpaper}). 

We envision an integrated system that centrally coordinates the formation of platoons. Trucks would connect to such a system via vehicle-to-infrastructure communication. The system continuously provides updated fuel-efficient routes and speed profiles to the connected trucks. These routes and speed profiles allow vehicles to meet on their journeys in order to form platoons. The computation of these routes and speed profiles takes various constraints such as arrival deadlines, speed-limits, and rest periods into account. 

The contribution of this paper is a method to efficiently rule out the majority of transport assignment pairs that cannot form a platoon due to their geographic or temporal separation. 
The elements of the significantly smaller set of candidate pairs is then treated one by one.

The computation of routes and coordinated speed profiles happens in several stages. The first stage is the route calculation that produces a route to the destination for each transport assignment based on the current position of the vehicle or the planned start point of the transport assignment in the future. It also computes a set of possible trajectories along that route. The next stage identifies for which pairs of transport assignments the associated vehicles can meet in the future in order to platoon based on the output of the route calculation. This is the input to an optimization routine that determines the platoons formed and the corresponding trajectories. These trajectories are sent to the individual vehicles which execute them. This process is frequently repeated in order to account for new or changed assignments, vehicles deviating from the planned trajectories, and updated traffic information. 
Comparing all pairs individually in order to find out which pairs can platoon is only feasible for a small number of transport assignments. 
On the other hand, the number of transport assignments increases when larger geographic regions or longer time horizons are considered in the planning process. In 2011 over 1.7 million heavy trucks were in use (\cite{hdv_statistics}) in the European Union, a number so large that even a fraction of these can be challenging to coordinate. 

A related problem to the one considered in this paper is to compute the collision of a large number of geometric objects, which has been considered in the field of computer graphics (\cite{collision_detection_survey, original_sweep_and_prune, original_sweep_and_prune2, i_collide, CUDAIntersection}) and to some extent in the area of interest management for distributed virtual environments (\cite{interest_management_survey}). 
The way this problem is tackled might also be relevant for other spatial, large-scale multi-agent systems where the possible interactions need to be identified in real time and where the number of actual interactions is small compared to the number of agent pairs. Examples of such systems are collision avoidance of (autonomous) mobile agents (\cite{tomlin_airspace}) and ride-sharing systems (\cite{ridesharing}). Similar to the application considered in this paper, these are large-scale spatially-distributed systems (\cite{spatial_networks}) that are enabled by the rapid development of the communication infrastructure. 

This paper is organized as follows. After having introduced the problem and related notation (Section~\ref{sec:problem_formulation}), we first abstract approaches that have been developed in computer graphics (Section~\ref{sec:culling}). We introduce the concept of features and classifiers that can indicate a pair of vehicles not being able to platoon on their routes. In Section~\ref{sec:features} we derive a family of features for the problem setting introduced in Section~\ref{sec:problem_formulation}. We demonstrate the method with a simulation study in Section~\ref{sec:simulations}.

\section{Problem Formulation}
\label{sec:problem_formulation}
 
We proceed with introducing the problem setup considered in this paper. We model the road network as a directed graph $\set{G}_\op{r} = (\set{N}_\op{r},\set{E}_\op{r})$ with nodes $\set{N}_\op{r}$ and edges $\set{E}_\op{r} \in \set{N}_\op{r} \times \set{N}_\op{r}$. Nodes correspond to intersections or endpoints in the road network and links correspond to road segments connecting these intersections. Each node in $\set{N}_\op{r}$ can be associated with a 2-D coordinate $\ve{P}: \set{N}_\op{r} \rightarrow \mathbb{R}^2$. 
Furthermore, we assume that the length of a road segment modeled by an edge equals the euclidean distance between the positions of the two nodes that comprise the edge. 

We have $K$ transport assignments and let $\set{N}_\op{c} = \{1, \dots, K\}$ be an index set of all assignments.
Each transport assignment consists of a start node $n^\op{S} \in \set{N}_\op{r}$ and a destination node $n^\op{D} \in \set{N}_\op{r}$. Furthermore, for each transport assignment, there is an earliest start time $t^\op{S}$ and a latest arrival time $t^\op{D}$.

The input to the vehicles are paths and speed profiles, one for each vehicle, that \textit{implement} the transport assignments. To that end, we define a trajectory (route and speed-profile) of a vehicle and the requirements for a trajectory to implement a transport assignment. 
\begin{defn}[Trajectory]
 A trajectory is a pair $(\ve{n},\ve{t})$. The route $\ve{n} = n[1], \dots, n[N]$ is a sequence of nodes in $\set{N}_\op{r}$ that describe a path in $\set{G}_\op{r}$. We refer to $\ve{n}$ as a route. The second element $\ve{t} = t[1], \dots, t[N]$ is a sequence of time instances such that $t[a+1] - t[a] \geq \|\ve{P}(n[a+1]) - \ve{P}(n[a])\|_2/v_\op{max}$ for $a = 1, \dots, N-1$, where $v_\op{max}$ is the maximum speed. The number of nodes $N$ may be different for different trajectories.
\end{defn}
We neglect that the speed on a link is restricted by the speed on the adjacent links and that the maximum speed $v_\op{max}$ depends on the link and on time. This is mainly for the ease of presentation but can as well be a reasonable simplification in the culling phase and can be accounted for in the routine that calculates the actual speed profiles. 
A trajectory $(\ve{n}, \ve{t})$ implements a transport assignment $(n^\op{S}, n^\op{D}, t^\op{S}, t^\op{D})$ if it starts after the start time at the start node of the transport assignment and arrives before the deadline at destination node of the transport assignment, i.e., if $n[1] = n^\op{S}$, $n[N] = n^\op{D}$, $t[1] \geq t^\op{S}$, $t[N] \leq t^\op{D}$.
We assume that the route $\ve{n}$ of the trajectory is given, typically the shortest path. The coordination is restricted to adapting the speed profile, i.e., the sequence $\ve{t}$.
We need to be able to test whether there are trajectories that platoon, i.e., partially coincide. To this end we calculate for each node in $\ve{n}$ a time interval.
The element in $\ve{t}$ that corresponds to the node in $\ve{n}$ lies in this interval if $\ve{t}$ belongs to a trajectory that implements the transport assignment. Only when the intervals for two transport assignments overlap at a common node, the possibility of implementing trajectories that platoon exists. We denote the sequence of lower bounds on the elements of $\ve{t}$ as $\underline{\ve{t}} = \underline{t}[1], \dots, \underline{t}[N]$ and the upper bounds as $\bar{\ve{t}} = \bar{t}[1], \dots, \bar{t}[N]$. They are computed for $a = 1, \dots, N$ as
\begin{align}
 \underline{t}[a] &= t^\op{S} + \sum\limits_{m = 1}^{a-1} \frac{\|\ve{P}(n[m+1]) - \ve{P}(n[m])\|_2}{v_\op{max}} \label{eq:lower_bound}\\
 \bar{t}[a] &= t^\op{D} - \sum\limits_{m = a}^{N-1} \frac{\|\ve{P}(n[m+1]) - \ve{P}(n[m])\|_2}{v_\op{max}}\label{eq:upper_bound}.
\end{align}

Next, we define a function that indicates whether platooning between two transport assignments is possible or not. This is the case if there is at least one common edge in the routes of the transport assignments where the time bounds of the two assignments overlap.
\begin{defn}[Coordination Function]\label{defi:coordination_function}
 The coordination function $\coordf: \set{N}_\op{c} \times \set{N}_\op{c} \rightarrow \{0, 1\}$ has the following properties. Let $\underline{\ve{t}}_i, \underline{\ve{t}}_j$ be lower bounds and $\bar{\ve{t}}_i, \bar{\ve{t}}_j$ be upper bounds on the node arrival times of transport assignments $i$ and $j$ according to \eqref{eq:lower_bound}, \eqref{eq:upper_bound}. Then it holds that $\coordf(i,j) = 1$, if there are indices $a, b$ such that $\ve{P}(n_i[a]) = \ve{P}(n_j[b])$ and $\ve{P}(n_i[a+1]) = \ve{P}(n_j[b+1])$, and $[\underline{t}_i[a], \bar{t}_i[a]] \cap [\underline{t}_j[b], \bar{t}_j[b]] \neq \emptyset$ and $[\underline{t}_i[a+1], \bar{t}_i[a+1]] \cap [\underline{t}_j[b+1], \bar{t}_j[b+1]] \neq \emptyset$. Otherwise $\coordf(i,j) = 0$.
\end{defn}
Comparing the routes and the time bounds in order to evaluate $\coordf$, is straightforward but computationally expensive. We refer to this as the \textit{exact algorithm}. 
The goal of this work is to find a scalable method for computing the set of all possible platoon pairs $\set{C} = \{(i,j) \in \set{N}_\op{c} \times \set{N}_\op{c}: \coordf(i,j) = 1\}$. Instead of iterating over all elements in $\set{N}_\op{c} \times \set{N}_\op{c}$ and using the exact algorithm, we propose to first efficiently compute an overapproximation $\hat{\set{C}} \supset \set{C}$ and then applying the exact algorithm.

\section{Culling}
\label{sec:culling}

The key idea of our approach is to extract features from the routes and time bounds ($\ve{n},\underline{\ve{t}}, \bar{\ve{t}}$) of the transport assignments to compute $\hat{\set{C}}$. These features can be more efficiently processed than $\ve{n}, \underline{\ve{t}}, \bar{\ve{t}}$. Features are designed in a way that no platooning opportunity in $\set{C}$ will be excluded from $\hat{\set{C}}$, so that $\set{C}$ can be computed from $\hat{\set{C}}$ using the exact algorithm. However, there might be some additional elements in $\hat{\set{C}}$ that do not actually correspond to platooning opportunities. We call these additional elements false-positives. The less false-positives there are in $\hat{\set{C}}$, the faster is the computation of $\set{C}$ from $\hat{\set{C}}$. This approach is inspired by a related problem of detecting which pairs of a large number of geometric objects intersect or collide.

We consider two types of features. These are \textit{interval features} and \textit{binary features}. Interval features map each object to an interval. The corresponding classifier indicates an intersection between two objects if the intervals generated by the objects overlap. There are algorithms (\cite{original_sweep_and_prune, original_sweep_and_prune2})
that can compute this classifier for all object pairs more efficiently than checking each pair individually, if the number of reported intersecting pairs is small.
Binary features map each object to a boolean value. The corresponding classifier indicates an intersection between two objects if the feature holds true for both objects. In Section~\ref{sec:features}, we derive appropriate features for the problem stated in Section~\ref{sec:problem_formulation}.

The classifiers are aggregated using boolean connectives. We formalize this in the remainder of the section. Let $\set{N}$ be a set of objects. We define a classifier as a function $c: \set{N} \times \set{N} \rightarrow \{0, 1\}$. If $c(i,j) = 0$, we call the combination of $c$ and $(i,j)$ a negative, and if $c(i,j) = 1$, we call it a positive.
Let $g: \set{N} \times \set{N} \rightarrow \{0, 1\}$ be the ground truth which can be computed by the exact algorithm. If for a pair $(i,j)$ we have $g(i,j) = 0$ and $c(i,j) = 1$, we call it a false-positive, and if $g(i,j) = 1$ and $c(i,j) = 0$, we call it a false-negative. Our aim is to design classifiers which yield no false negatives for all elements of $\set{N} \times \set{N}$ and few false-positives that have to be processed by the exact algorithm in addition to the true-positives.  

We can identify two types of basic classifiers that are combined in a specific way in order to achieve the above objective.
A classifier $c$ is \textit{required} if $\neg c(i,j) \Rightarrow \neg g(i,j)$ for all $i, j \in \set{N} \times \set{N}$. 
In some cases, we have to take into account a set of classifiers to conclude that $g$ does not hold.
A set of classifiers $\set{D}$ is \textit{required} if $\neg \bigvee\limits_{c \in \set{D}} c(i,j) \Rightarrow \neg g(i,j)$ for all $i, j \in \set{N} \times \set{N}$. 
It is straightforward to construct a required classifier from a required set of classifiers.
\begin{prop}\label{prop:required_set}
If a set $\set{D}$ of classifiers is required, then $\bigvee\limits_{c \in \set{D}} c$ is a required classifier.
\end{prop}
We can combine two required classifiers into one required classifier that performs no worse than any of the required classifiers it is combined of.
\begin{prop}\label{prop:combination_of_two_classifiers}
 If $c_1$ and $c_2$ are required classifiers, then $c_{12} := c_1 \wedge c_2$ is a required classifier. Let $\bar{\set{E}}_{12} = \{(i,j) \in \set{N} \times \set{N}: c_{12}(i,j) = 0\}$ be the set of negatives of $c_{12}$ and let $\bar{\set{E}}_1$, $\bar{\set{E}}_2$ be the set of negatives for $c_1$ and $c_2$ respectively. Then $\bar{\set{E}}_1 \subseteq \bar{\set{E}}_{12}$ and $\bar{\set{E}}_2 \subseteq \bar{\set{E}}_{12}$.
\end{prop}
\begin{pf}
 For $c_{12}$ to be required, we need to show that $\neg c_{12}(i,j) \Rightarrow \neg g(i,j)$ for all $i, j \in \set{N} \times \set{N}$. 
 We have
$
  (\neg c_1 \Rightarrow \neg g) \wedge (\neg c_2 \Rightarrow \neg g) 
  = (c_1 \vee \neg c_1 \wedge \neg g) \wedge (c_2 \vee \neg c_2 \wedge \neg g) 
  = c_1 \wedge c_2 \vee \neg g \wedge(\neg c_1 \wedge \neg c_2 \vee \neg c_1 \wedge c_2 \vee c_1 \wedge \neg c_2) 
  = c_1 \wedge c_2 \vee \neg g \wedge(\neg c_1 \vee \neg c_2)
  = c_1  \wedge c_2 \vee \neg g \wedge \neg (c_1 \wedge c_2)
  = \neg (c_1 \wedge c_2) \Rightarrow \neg g
  = \neg c_{12} \Rightarrow \neg g
$.
Let $(i,j) \in \bar{\set{E}}_1$. Then from the definition of $\bar{\set{E}}_1$ we have that $c_1(i,j) = 0$. We have that $c_{12}(i,j) = c_1(i,j) \wedge c_2(i,j) = 0 \wedge c_2(i,j) = 0$. It follows from the definition of $\bar{\set{E}}_{12}$ that $(i,j) \in \bar{\set{E}}_{12}$. Similarly, we see that any element of $\bar{\set{E}}_1$ is an element of $\bar{\set{E}}_{12}$.
\end{pf}
In this manner, we can combine as many required classifiers as we want and have at our disposal. With each classifier we add, we potentially decrease the set of remaining candidates that need to be checked by the exact algorithm. There is a trade-off between doing more work to evaluate more classifiers and less instances which have to be processed by the exact algorithm (\cite{CUDAIntersection}).

\section{Features and Classifiers}
\label{sec:features}

In order to apply the results from Section~\ref{sec:culling}, we need to specify appropriate features and classifiers based on these features for the problem stated in Section~\ref{sec:problem_formulation}. Once we know how to compute appropriate features that yield required classifiers or required sets of classifiers, we can use the results from Section~\ref{sec:culling} to execute the culling phase. The remaining candidate pairs are passed on to the exact algorithm to compute $\set{C}$. Hence, we will derive a selection of features and corresponding classifiers in this section. In Section~\ref{sec:simulations}, we will demonstrate these classifiers and combinations of them in a simulation example. 

\begin{figure}[t]
\begin{center}
\def\svgwidth{.7\columnwidth} 
 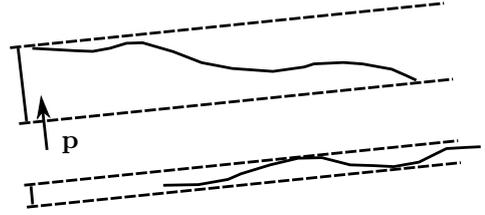
 \caption{Illustration of the projection feature. It shows how the two routes (solid lines) are projected onto a line in the direction of the vector $\ve{p}$. The borders of the intervals are indicated with dashed lines. For illustration purposes the third dimension is omitted here.}
\label{fig:projection_classifier}
\end{center}
\end{figure}

The first feature projects the possible trajectories on a line which yields an interval.
Formally, we define this feature as follows. 
\begin{defn}\label{defi:bounding_box_feature}
Let $\ve{p} \in \mathbb{R}^3$ be a three dimensional vector which defines the orientation of the line which the trajectories are projected onto. Then the associated interval feature is defined as
\begin{equation}
 \set{I} = [\min\limits_{\ve{v} \in \set{R}} (\ve{p}^\transp \ve{v}), \max\limits_{\ve{v} \in \set{R}} (\ve{p}^\transp \ve{v})]
 \label{eq:projection_interval}
\end{equation}
with
\begin{equation}
\begin{split}
\set{R} = \left\{
\begin{bmatrix}\ve{P}(n[1])\\ \underline{t}[1]\end{bmatrix}, \dots, \begin{bmatrix}\ve{P}(n[N])\\ \underline{t}[N]\end{bmatrix},\right. \\ \left.
\begin{bmatrix}\ve{P}(n[1])\\ \bar{t}[1]\end{bmatrix}, \dots, \begin{bmatrix}\ve{P}(n[N])\\ \bar{t}[N]\end{bmatrix}
\right\}.\label{eq:def_R}
\end{split}
\end{equation}
\end{defn}
This feature is illustrated in Figure~\ref{fig:projection_classifier}. The projection vector $\ve{p}$ is a design choice. 
Proposition~\ref{prop:combination_of_two_classifiers} allows us to combine arbitrarily many classifiers based on this kind of feature with different $\ve{p}$.

Next, we establish that if for a pair of transport assignments the intervals do not overlap the coordination function is equal to zero. This allows us to define a required feature based on the overlap between these intervals. 
\begin{prop}\label{theo:projection_classifier}
 Let $(i,j)$ refer to a pair of transport assignments. Let $\set{I}_i, \set{I}_j$ be the interval features according to \eqref{eq:projection_interval} for the two transport assignments. 
 Then $\set{I}_i \cap \set{I}_j = \emptyset \Rightarrow \coordf(i,j) = 0$.
\end{prop}
\begin{pf}
 According to Definition~\ref{defi:coordination_function}, $\coordf(i,j) = 1$ implies that there must be indices $a, b$ such that $\ve{P}(n_i[a]) = \ve{P}(n_j[b])$ and $[\underline{t}_i[a], \bar{t}_i[a]] \cap [\underline{t}_j[b], \bar{t}_j[b]] \neq \emptyset$, where $\ve{n}_i, \underline{\ve{t}}_i, \bar{\ve{t}}_i$ and $\ve{n}_j,\underline{\ve{t}}_j, \bar{\ve{t}}_j$ are the node sequences and time bounds of transport assignment $i, j$ respectively. We have
 \begin{equation*}
  [\underline{t}_i[a], \bar{t}_i[a]] \cap [\underline{t}_j[b], \bar{t}_j[b]] \neq \emptyset
  \Leftrightarrow
  \underline{t}_i[a] \leq \bar{t}_j[b] \wedge \underline{t}_j[b] \leq \bar{t}_i[a].
 \end{equation*}
 Let $\ve{p} = [p_1, p_2, p_3]^\transp$, $\ve{P} = \ve{P}(n_i[a]) = \ve{P}(n_j[b])$, and $P^0 = [p_1, p_2] \ve{P}$. We have
\small
 \begin{align*}
  &\underline{t}_i[a] \leq \bar{t}_j[b] \wedge \underline{t}_j[b] \leq \bar{t}_i[a]
  \\
  \Rightarrow
  &\min ( p_3 \underline{t}_i[a], p_3 \bar{t}_i[a]) \leq \max (p_3 \underline{t}_j[b], p_3 \bar{t}_j[b])
  \\
  \Rightarrow
  &\min ( p_3 \underline{t}_i[a] + P^0, p_3 \bar{t}_i[a] + P^0) 
  \leq \max (p_3 \underline{t}_j[b] + P^0, p_3 \bar{t}_j[b] + P^0)
  \\
  \Rightarrow
  &\min \left(\ve{p}^\transp \begin{bmatrix}\ve{P}\\ \underline{t}_i[a]\end{bmatrix}, 
              \ve{p}^\transp \begin{bmatrix}\ve{P}\\ \bar{t}_i[a]\end{bmatrix}\right)
  \leq
   \max  \left(\ve{p}^\transp \begin{bmatrix}\ve{P}\\ \underline{t}_j[b]\end{bmatrix}, 
               \ve{p}^\transp \begin{bmatrix}\ve{P}\\ \bar{t}_j[b]\end{bmatrix}\right)
  \\
  \Rightarrow
  &\min\limits_{\ve{v} \in \set{R}_i} (\ve{p}^\transp \ve{v})
   \leq
   \max\limits_{\ve{v} \in \set{R}_j} (\ve{p}^\transp \ve{v})
   ,
 \end{align*}
\normalsize
with $\set{R}_i, \set{R}_j$ as in \eqref{eq:def_R} for transport assignment $i, j$, respectively.
Similarly, by swapping $i$ and $j$, we can show that the conditions of the proposition imply that
\begin{align*}
 \min\limits_{\ve{v} \in \set{R}_j} (\ve{p}^\transp \ve{v})
   \leq
   \max\limits_{\ve{v} \in \set{R}_i} (\ve{p}^\transp \ve{v}).
\end{align*}
The above two conditions combined imply that $\set{I}_i \cap \set{I}_j \neq \emptyset$. Thus $\coordf = 1 \Rightarrow \set{I}_i \cap \set{I}_j \neq \emptyset$ or equivalently $\set{I}_i \cap \set{I}_j = \emptyset \Rightarrow \coordf = 0$.

\end{pf}
 
Next, we will introduce a binary feature that leads to a required classifier. This feature is based on the orientations of the individual links in a route. It will only be useful if all segments in a route point approximately from start to goal location. Later on, we will address the problem of outliers. Here, we derive a set of required classifiers each based on a binary feature from the orientation. 
 The orientation $\Theta(n_1,n_2) \in [0, 2\pi]$ of an edge $(n_1, n_2) \in \set{E}_\op{r}$ is the angle in polar coordinates of the vector $\ve{P}(n_2) - \ve{P}(n_1)$.
We choose a partition of the interval $[0, 2\pi]$. Each element of the partition is related to one binary feature which holds true if the orientation of at least one edge in the route falls in the range of that element. When two routes overlap there must be at least one edge that has the same orientation.
\begin{prop}\label{prop:orientation_classifier}
 Let $(i,j)$ refer to the pair of transport assignments. Let $\set{P}$ be a partition of $[0, 2\pi]$. If there is no element $I \in \set{P}$ and edges in the routes of the transport assignments $(n_i[a], n_i[a+1])$, $(n_j[b], n_j[b+1])$ such that $\Theta(n_i[a], n_i[a+1]) \in I$ and $\Theta(n_j[b], n_j[b+1]) \in I$, then $\coordf(i,j) = 0$.
\end{prop}
\begin{pf}
 According to Definition~\ref{defi:coordination_function}, $\coordf(i,j) = 1$ implies that there must be indices $a, b$ such that $\ve{P}(n_i[a]) = \ve{P}(n_j[b])$ and $\ve{P}(n_i[a+1]) = \ve{P}(n_j[b+1])$, where $n_i, n_j$ are the node sequences of transport assignment $i, j$ respectively. For these it holds that $\Theta(n_i[a],n_i[a+1]) = \Theta(n_j[b],n_j[b+1])$.
 Since $\set{P}$ is a partition of $[0, 2\pi]$ and $\Theta(n_i[a],n_i[a+1]) \in [0, 2\pi]$, there must be $I \in \set{P}$ with $\Theta(n_i[a],n_i[a+1]) \in \set{I}$. Since $\Theta(n_j[b],n_j[b+1]) = \Theta(n_i[a],n_i[a+1])$, it follows that also $\Theta(n_j[b],n_j[b+1]) \in \set{I}$. The proof follows from contradiction. 
\end{pf}

Next, we discuss how we can make this classifier based on orientation more efficient if we can disregard routes that overlap only over a short distance. 
Apart from the direct reduction in true positives, this approach will also reduce the false-positive rate of the classifiers, since some outlier route edges can be disregarded. 

In order to cover the notion that there must be a minimum overlap in routes to be considered, we extend the definition of the coordination function (Definition~\ref{defi:coordination_function}).
\begin{defn}[Min. Distance Coordination Function]\label{defi:min_dist_coord_fun}
 \hspace{1mm}\\
 A coordination function $\coordf: \set{N}_\op{c} \times \set{N}_\op{c} \rightarrow \{0,1\}$  according to Definition~\ref{defi:coordination_function} requires minimum distance $l_{\min}$ if the following properties hold: if for a pair $(i,j)$ we have $\coordf(i,j) = 1$, there must be a set of pairs of indices $\set{A}$ such that for all $(a,b) \in \set{A}$ it holds that $\ve{P}(n_i[a]) = \ve{P}(n_j[b])$ and $\ve{P}(n_i[a+1]) = \ve{P}(n_j[b+1])$, and $[\underline{t}_i[a], \bar{t}_i[a]] \cap [\underline{t}_j[b], \bar{t}_j[b]] \neq \emptyset$ and $[\underline{t}_i[a+1], \bar{t}_i[a+1]] \cap [\underline{t}_j[b+1], \bar{t}_j[b+1]] \neq \emptyset$. Furthermore we require $\sum\limits_{(a,b) \in \set{A}} \|\ve{P}(n_i[a]) - \ve{P}(n_i[a+1])\|_2 \geq l_{\min}$.
\end{defn}
We adapt the orientation-based classifier (Proposition~\ref{prop:orientation_classifier}) to exclude links of a total length less than $l_{\min}$.
The approach is to calculate the fraction of route length that lies in each element of the partition. We can ignore the intersection with some elements of the partition as long as the lengths of the links whose orientation is contained in these elements sums up to a value less than  $l_{\min}/2$. 
\begin{prop}\label{prop:orientation_classifier_l_min}
 Let $(i,j)$ refer to a pair of transport assignments. Let $\set{P}$ be a partition of $[0, 2\pi]$. Let $\set{I}_i \subseteq \set{P}$ and let $\bar{\set{E}}_i \subseteq \set{E}_i$, where $\set{E}_i = \{(n_i[a],n_i[a+1]): a \in \{1, \dots, N_i\}\}$, such that for all $e \in \bar{\set{E}}_i$, it holds that there exists $I \in \set{I}_i$ with $\Theta(e) \in I$ and we have $\sum\limits_{(n_1,n_2) \in \set{E}_i \setminus \bar{\set{E}}_i} \|\ve{P}(n_1) - \ve{P}(n_2)\|_2 < l_{\min}/2$. Similarly, by replacing $i$ by $j$, we define $\set{I}_j$ for transport assignment $j$. If $\set{I}_i \cap \set{I}_j = \emptyset$, then $\coordf(i,j) = 0$ with $\coordf$ according to Definition~\ref{defi:min_dist_coord_fun}.
\end{prop}
\begin{pf}
 If $\coordf(i,j) = 1$, then we have a set of pairs of indices $\set{A}$ such that for all $(a,b) \in \set{A}$ it holds that $\ve{P}(n_i[a]) = \ve{P}(n_j[b])$ and $\ve{P}(n_i[a+1]) = \ve{P}(n_j[b+1])$. Thus it also holds that $\Theta(n_i[a], n_i[a+1]) = \Theta(n_j[b], n_j[b+1])$. Since $\set{P}$ is a partition of the image of $\Theta(\cdot)$, there is exactly one element $I \in \set{P}$ with $\Theta(n_i[a], n_i[a+1]) \in I$ and since  $\Theta(n_i[a], n_i[a+1]) = \Theta(n_j[b], n_j[b+1])$, we have $\Theta(n_i[a], n_i[a+1]) \in I \Leftrightarrow \Theta(n_j[b], n_j[b+1]) \in I$. 
 Furthermore, we have from Definition~\ref{defi:min_dist_coord_fun} that $\sum\limits_{(a,b) \in \set{A}} \|\ve{P}(n_i[a]) - \ve{P}(n_i[a+1])\|_2 \geq l_{\min}$. 
 
 Let $\bar{\set{A}}_i$ be a set of the indices of the head nodes of edges in $(\set{E}_i \cap \set{E}_j) \setminus \bar{\set{E}}_i$ paired with the corresponding indices in route $j$, with $\set{E}_i, \set{E}_j, \bar{\set{E}}_i$ as defined in the proposition. These are the pairs of indices of the edges in the common part of the route that are ignored in transport assignment $i$. Similarly, let $\bar{\set{A}}_j$ be the index pairs that are excluded due to transport assignment $j$. We need to show now that $\set{A}$ is not empty without the pairs in $\bar{\set{A}}_i$ and $\bar{\set{A}}_j$, or in other words, that even if the features for either route ignore up to $l_{\min}/2$ of the common part of the route, there are still edges left that let the set of classifiers indicate that the routes intersect.
 We have from the assumptions made in the proposition $\sum\limits_{(a,b) \in \bar{\set{A}}_i} \|\ve{P}(n_i[a]) - \ve{P}(n_i[a+1])\|_2 < l_{\min}/2$, $\sum\limits_{(a,b) \in \bar{\set{A}}_j} \|\ve{P}(n_i[a]) - \ve{P}(n_i[a+1])\|_2 < l_{\min}/2$, and from Definition~\ref{defi:min_dist_coord_fun} that $\sum\limits_{(a,b) \in \set{A}} \|\ve{P}(n_i[a]) - \ve{P}(n_i[a+1])\|_2 \geq l_{\min}$. Thus $\sum\limits_{(a,b) \in \set{A} \setminus (\bar{\set{A}}_j \cup \bar{\set{A}}_j)} \|\ve{P}(n_i[a]) - \ve{P}(n_i[a+1])\|_2 > 0$ and since this is a sum over positive elements, we deduce that $\set{A} \setminus (\bar{\set{A}}_j \cup \bar{\set{A}}_j) \neq \emptyset$. But then there is $I \in \set{P}$ and $(a,b) \in \set{A} \setminus (\bar{\set{A}}_j \cup \bar{\set{A}}_j)$ such that $\Theta(n_i[a], n_i[a+1]) = \Theta(n_j[b], n_j[b+1]) \in I$ and thus $\set{I}_i \cap \set{I}_j \neq \emptyset$. By contraposition it follows that $\set{I}_i \cap \set{I}_j = \emptyset \implies \coordf(i,j) = 0$.
\end{pf}
It is possible to combine various classifiers as defined in Propositions~\ref{theo:projection_classifier} and \ref{prop:orientation_classifier_l_min} in various ways according to Propositions~\ref{prop:required_set} and \ref{prop:combination_of_two_classifiers} in Section~\ref{sec:culling}.  

\section{Simulations}
\label{sec:simulations}

In this section, we demonstrate in a realistic scenario the method derived in this paper. We demonstrate that the application of 6 classifiers can rule out 99\,\% of the transport assignment pairs, leaving only 1\;\% for the computationally expensive exact algorithm.
The simulation setup is as follows. The start and goal locations are sampled randomly with probability proportional to an estimate of the population density in the year 2000 (\cite{population_density}). We limit the area to a large part of Europe.
We calculate shortest routes with the Open Source Routing Machine (\cite{project_osrm}). If the route is longer than 400 kilometers, a subsection of 400 kilometers of the route is randomly selected. 
The maximum speed is $v_{\max} = 80\,$km/h. We set the start times $t^\op{S}$  of half the assignments to 0 and sample the start times of the remaining assignments uniformly in an interval of 0 to 24\,h. The first half is to account for assignments that are currently on the road while the other half is to account for assignments that are scheduled to depart later. The deadlines $t^\op{D}$ are set in such a way that the interval $\bar{t}[a] - \underline{t}[a] = 0.5\,h$ where $a$ is any valid index. We consider the minimum length that two assignments have to overlap to be considered for platooning, $l_{\min}$, to be 20\,km.

We implemented all features and corresponding classifiers that are described in Section~\ref{sec:features}, i.e., interval projection (Proposition~\ref{theo:projection_classifier}) and minimum distance orientation partition (Proposition~\ref{prop:orientation_classifier_l_min}). Note that Proposition~\ref{prop:orientation_classifier} is a special case of Proposition~\ref{theo:projection_classifier} with $l_{\min} = 0$. For interval projection we tested vectors of the form 
 $\begin{bmatrix} 1 , 0 , 0 \end{bmatrix}^\transp$,
 $\begin{bmatrix} 0 , 1 , 0 \end{bmatrix}^\transp$,
 $\begin{bmatrix} 0 , 0 , 1 \end{bmatrix}^\transp$,
 $\begin{bmatrix} 1 , 1 , 0 \end{bmatrix}^\transp$,
 $\begin{bmatrix} -1 , 1 , 0 \end{bmatrix}^\transp$,
 $\begin{bmatrix} -\cos(\alpha), \frac{-\sin(\alpha)}{\cos(0.278\pi )} , \frac{v_{\max} 180^\circ}{6371\pi} \end{bmatrix}^\transp$,
with $\alpha = 0, \pi/4, \dots, 7\pi/4$. The position $\ve{P}$ is expressed here as latitude and longitude and measured in degrees. The vectors parametrized by $\alpha$ are approximately orthogonal to a trajectory at maximum speed at the latitude of 50 degrees with heading angle $\alpha$ and should work well for trajectory pairs that have similar orientation, that cover the same area, and that are only separated by a small time margin. We will refer to the corresponding classifiers in the following discussion as $c_{100}, c_{010}, c_{001}, c_{110}, c_{-110}, c_{\alpha 0}, \dots, c_{\alpha 7}$ respectively. For the orientation-based classifier we use 100 equally sized cells to partition $[0, 2\pi]$. For each cell the fraction of the route distance that falls in this cell is computed. Matches up to $l_{\min}/2$ starting in ascending order of route distance contained in the cells are excluded. We will refer to this classifier as $c_{\op{o}}$.

\begin{figure}[t]
  \begin{center}
 \input{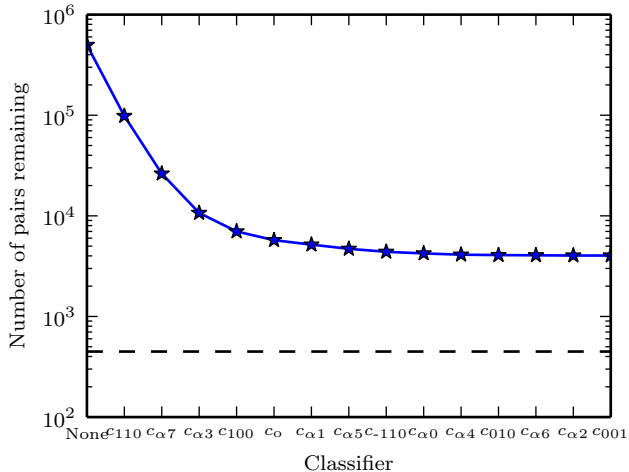}
 \caption{This plot shows the number of remaining pairs when the classifiers are consecutively applied from left to right. The order the classifiers are chosen in a way that each stage removes as many pairs as possible. The classifier applied at each stage is indicated on the horizontal axis. The dashed line shows the ground truth from the exact algorithm.}
\label{fig:classifier_order}
\end{center}
\end{figure}

We test $K = 1000$ transport assignments. All classifiers are evaluated in parallel. Then the sequence of classifiers that filters the most assignments at every stage is computed. The number of positives for each classifier is listed in Table~\ref{tab:number_of_matches}. Figure~\ref{fig:classifier_order} shows the number of remaining pairs at each stage, the ground truth, and the sequence of classifiers for this sample. The optimization of the classifier order would typically be done when the system is designed and is to some extent specific for the exact problem setting. In a running platoon coordination system the order in which classifiers are applied would remain fixed.

We can see in Figure~\ref{fig:classifier_order} that two classifiers, $c_{110}$ and $c_{\alpha 7}$, combined are able to reduce the number of pairs by one order of magnitude. The first classifier, $c_{110}$, only takes into account longitude and latitude of the routes. The second one, $c_{\alpha 7}$ is orthogonal to the first one, $c_{110}$, in the plane but also takes into account timing. The third classifier, $c_{\alpha 3}$, is also of the projection type that is able to identify that a pair of assignments cannot platoon if they are geographically close but differ in timing, and it covers the opposite orientation compared to the previous classifier. The fourth classifier, $c_{100}$, covers a third direction in the plane. It is interesting to see that the fifth classifier, $c_{\op{0}}$, is the orientation-based classifier. Alone it performs much worse than the other classifiers as can be seen in Table~\ref{tab:number_of_matches}. Two transport assignments that take the same route in opposite directions and that ``meet'' on the way are impossible to identify as a negative with the projection based classifiers. The orientation-based classifier might be able to achieve that. The classifier that only takes into account start and arrival time, $c_{001}$, is selected last, since most cases it rules out are already covered by the classifiers $c_{\alpha 0}, \dots, c_{\alpha 7}$ and since half the assignments start at the same time. We see that the benefit from adding more classifiers diminishes quickly as classifiers are added. All classifiers combined can reduce the number of pairs by two orders of magnitude and get within one order of magnitude from the ground truth. The false-positives are mostly very curvy routes that intersect geographically and are separated little in time in the area of the intersection. To be able to correctly identify such pairs as negatives is often not possible with the features presented in this paper. We get fairly consistent results for different runs of the simulation that are omitted here owing to space constraints. 

\begin{table}[t]
\begin{center}
\begin{tabular}{lr|lr|lr}
 None & 499,500 & $c_{-110}$ & 108,403 & $c_{\alpha 4}$ & 134,019 \\
 $c_{100}$ & 104,380 & $c_{\alpha 0}$ & 129,282 & $c_{\alpha 5}$ & 107,287 \\
 $c_{010}$ & 101,542 & $c_{\alpha 1}$ & 103,240 & $c_{\alpha 6}$ & 105,883 \\
 $c_{001}$ & 208,896 & $c_{\alpha 2}$ & 103,453 & $c_{\alpha 7}$ & 109,934 \\
 $c_{110}$ & 98,343 & $c_{\alpha 3}$ & 109,626 & $c_{\op{o}}$ & 453,246 \\
\end{tabular}
\end{center}
\caption{Number of positives for different classifiers.}
\label{tab:number_of_matches}
\end{table}

\section{Conclusions and Future Work}

We presented a method that can significantly reduce the computational effort of centrally coordinating truck platooning over large geographic areas and time intervals. With this framework we can combine different classifiers to efficiently cull the pairs of transport assignments before they are passed on to an exact algorithm that checks which transport assignments can platoon. We developed three types of classifiers and demonstrated their potential in simulation. This approach might also be useful for dynamically computing the interaction network of spatially distributed multi-agent systems. 

There are various directions for future work. While this work focuses on algorithmic efficiency, it would be interesting to tune the implementation for best performance. 
Furthermore, we would like to gain more insight from simulations and from theoretical analysis on how different parameters affect the effectiveness of the approach. It would be, for instance, desirable to better understand which classifiers should be chosen in which order based on the way transport assignments are generated.

\bibliography{citations}
\bibliographystyle{ifacconf}

\end{document}